\documentclass[twocolumn,preprintnumbers,superscriptaddress,nofootinbib,aps,prl,floatfix]{revtex4}

\usepackage{amsmath,amssymb}
\usepackage{dsfont}
\usepackage{graphicx} %loads the graphicx.sty package 
\usepackage{epstopdf} %loads the epstopdf.sty package 
\usepackage{slashed}
\usepackage{subfigure}
\usepackage{color}
\usepackage{multirow}
\usepackage{hyperref}

\usepackage{footmisc}

\newcommand{\definmath}[2] {\def#1{\ifmmode#2\else$#2$\fi}}

\newcommand{\GeV}{\,\mathrm{GeV}}

\definmath{\pslash}{\!\not\! p}
\definmath{\pt}{p_{\rm T}}
\definmath{\genericT}{{\rm T}}
\definmath{\ptmiss}{\slashed{p}_\genericT}
\definmath{\mptvec}{\slashed{\vec{p}}_\genericT}

\definmath{\mt}{m_{\rm T}}
\definmath{\mttwo}{{m_{\rm T2}}}
\definmath{\mboundfull}{{m_{\tau\tau}^{\rm Higgs-bound}}}
\definmath{\mboundgeneral}{{m_{\tau\tau}^>}}

\definmath{\ptbbbar}{p_{{\rm T},b\bar b}}
\definmath{\invfb}{\mathrm{fb}^{-1}}
\definmath{\invab}{\mathrm{ab}^{-1}}
\definmath{\ttbar}{{t\bar{t}}}
\definmath{\sumEt}{{\Sigma E_{\rm T}}}
\newcommand{\paper}{letter}
\newcommand{\mysection}[1]{\section{#1}}
\newcommand{\mysubsection}[1]{\vspace{-5mm}\subsection{#1}\vspace{-2mm}}

\usepackage{array}
\newcolumntype{L}[1]{>{\raggedright\let\newline\\\arraybackslash\hspace{0pt}}m{#1}}
\newcolumntype{C}[1]{>{\centering\let\newline\\\arraybackslash\hspace{0pt}}m{#1}}
\newcolumntype{R}[1]{>{\raggedleft\let\newline\\\arraybackslash\hspace{0pt}}m{#1}}

\begin{document}  

\author{Alan J. Barr} \email{a.barr@physics.ox.ac.uk}
\affiliation{Denys Wilkinson Building, Department of Physics,\\ Keble Road,
  Oxford, OX1 3RH, United Kingdom\\[0.1cm]}
\author{Matthew J. Dolan} \email{m.j.dolan@durham.ac.uk}
\affiliation{Institute for Particle Physics Phenomenology, Department
  of Physics,\\Durham University, DH1 3LE, United Kingdom\\[0.1cm]}
\author{Christoph Englert} \email{christoph.englert@glasgow.ac.uk}
\affiliation{SUPA, School of Physics and Astronomy, University of
  Glasgow,\\Glasgow, G12 8QQ, United Kingdom\\[0.1cm]}
\author{Michael Spannowsky} \email{michael.spannowsky@durham.ac.uk}
\affiliation{Institute for Particle Physics Phenomenology, Department
  of Physics,\\Durham University, DH1 3LE, United Kingdom\\[0.1cm]}

\preprint{IPPP/13/74, DCPT/13/148}
 
\title{Di-Higgs final states augMT2ed -- \\
selecting $hh$ events at the high luminosity LHC
}

%%%%%%%%%%%%%%%%%%%%%%%%%%%%%%%%%%%%%%
\begin{abstract} 
%%%%%%%%%%%%%%%%%%%%%%%%%%%%%%%%%%%%%%
  Higgs boson self-interactions can be investigated via di-Higgs
  ($pp\to hh+X$) production at the LHC.
  With a small ${\cal{O}}(30)$~fb Standard Model production cross
  section, and a large \ttbar{} background, this measurement has been
  considered challenging, even at a luminosity-upgraded LHC.  We
  demonstrate that by using simple kinematic bounding variables, of
  the sort already employed in existing LHC searches, the dominant
  \ttbar{} background can be largely eliminated. Simulations of the
  signal and the dominant background demonstrate the prospect for
  measurement of the di-Higgs production cross section at the $30\%$
  level using $3\,\invab$ of integrated luminosity at a
  high-luminosity LHC. This corresponds to a Higgs self-coupling
  determination with $60\%$ accuracy in the $b\bar b \tau^+\tau^-$
  mode, with potential for further improvements from e.g. subjet
  technologies and from additional di-Higgs decay channels.
%
%%%%%%%%%%%%%%%%%%%%%%%%%%%%%%%%%%%%%%
\end{abstract}   
%%%%%%%%%%%%%%%%%%%%%%%%%%%%%%%%%%%%%%
\maketitle 

%%%%%%%%%%%%%%%%%%%%%%%%%%%%%%%%%%%%%%
\mysection{Introduction}
%%%%%%%%%%%%%%%%%%%%%%%%%%%%%%%%%%%%%%
After a particle consistent with the Standard Model (SM) Higgs boson
has been discovered at the LHC \cite{Aad:2012tfa,Chatrchyan:2012ufa},
we have the final irrefutable experimental evidence of the realisation
of a Higgs mechanism in nature
\cite{Higgs:1964ia,Higgs:1964pj,Guralnik:1964eu,Englert:1964et}. This
discovery alone, however, does not provide us the full details of this
symmetry breaking sector. In particular, we do not have any additional
information other than the existence of a (local) symmetry-breaking
minimum and the Higgs potential's curvature at this point in field
space. These are rather generic properties of symmetry breaking
potentials which can easily be reconciled with more complex scenarios
of electroweak symmetry breaking.  These typically exhibit a
significantly different form of the Higgs self-interaction from the
SM\footnote{For example in scenarios in which the electroweak symmetry
  is broken radiatively, we typically encounter Coleman-Weinberg type
  potentials~\cite{Coleman:1973jx} which exhibit an infinite power
  series in the Higgs field with model-dependent expansion
  coefficients.}, and to obtain a better understanding of how
electroweak symmetry breaking comes about, we need to find a way to
discriminate between these different realisations.

The only direct way to provide a satisfying discrimination between the
SM symmetry breaking sector and more complicated realisations is
probing higher order terms of the Higgs potential directly. In
practice this means studying multi-Higgs final states and inferring
the relevant couplings from data.  The size of the cross sections at
the LHC and future colliders effectively limits such a program to the
investigation of the trilinear Higgs
coupling~$\lambda$~\cite{Plehn:2005nk}. In the SM, $\lambda$ is a
function of the Higgs mass $m_h$ and the quartic Higgs interaction
$\eta$,
\begin{multline}\nonumber
  %\label{eq:higgspot}
  V(H^\dagger H) = 
    \mu^2 H^\dagger H + \eta (H^\dagger H)^2 \\
    \rightarrow {1\over 2} m_h^2h^2 + \sqrt{ {\eta\over 2}}m_h h^3
    + {\eta\over 4}h^4\,,
\end{multline}
where we have expanded the potential around the non-zero Higgs vacuum
expectation value in unitary gauge in the second line which yields
$\lambda_{\text{SM}}=m_h\sqrt{\eta/2}$.

The effort of phenomenologically reconstructing the trilinear Higgs
coupling is based on di-Higgs production $pp\to hh+X$
\cite{Glover:1987nx,Dicus:1987ic,Plehn:1996wb,Djouadi:1999rca,deFlorian:2013uza,Grigo}
and dates back more than a decade~\cite{Baur:2002qd}, but in the light
of the recent Higgs discovery it has gained new momentum
\cite{Dolan:2012rv,Papaefstathiou:2012qe,Dolan:2012ac,Baglio:2012np,Goertz:2013kp,Cao:2013si,Gupta:2013zza,Grigo:2013rya,Nhung:2013lpa,Ellwanger:2013ova}.
Probably the most promising approach to infer the trilinear coupling
which has been proposed so far is via the $hh\to b\bar b\tau^+\tau^-$
channel at the LHC, using boosted techniques
\cite{Butterworth:2008iy,Plehn:2009rk} as reported first in the
hadron-level analysis of Ref.~\cite{Dolan:2012rv}.  That analysis was
conservative in the sense that it did not employ selection criteria
based on missing transverse momentum, which have the potential to
reduce the most challenging $t\bar t $ backgrounds.

In the present \paper{} we complement the analysis of
Ref.~\cite{Dolan:2012rv} along these lines and also address the
question of the extent to which a successful analysis of the di-Higgs
final state will depend on the overall Higgs boost.  We concentrate on
the $b\bar b\tau^+\tau^-$ mode,
\begin{equation}
  \label{eq:hhbbtautau}
  pp \;\rightarrow  \;hh +X\; \rightarrow 
  \; ( b + \bar{b} ) + ( \tau^+ + \tau^- ) + X,
\end{equation}
for which the $\ttbar$ background process
\begin{multline}
  \label{eq:ttbarbg}
  pp  \; \rightarrow  \; \ttbar +X  \; \rightarrow  
  \;(b + W^+) + (\bar b + W^-) + X \\ \rightarrow 
  (b + \tau^+ + \nu_\tau) + (\bar b + \tau^- + \bar \nu_\tau)  +X 
\end{multline}
dominates.  We use kinematical properties of the decay of
Eq.~\eqref{eq:ttbarbg} to greatly reduce the \ttbar{} background.

While we focus on the $b\bar b\tau^+\tau^-$ mode in this \paper{}, we
note that variants of the technique would be applicable to a broader
range of di-Higgs decay modes, particularly others also involving the
$h \rightarrow b\bar{b}$ and $h\to W^+W^-$ decays, which have the
largest branching ratios for a 125~GeV Standard Model Higgs boson.

%%%%%%%%%%%%%%%%%%%%%%%%%%%%%%%%%%%%%%
\mysection{Kinematic bounding variables}
%%%%%%%%%%%%%%%%%%%%%%%%%%%%%%%%%%%%%%
The dominant \ttbar{} background can be reduced by using the \mttwo{}
variable, sometimes called the `stransverse
mass'~\cite{Lester:1999tx,Barr:2003rg}.  This mass-bound variable was
designed for the case where a pair of equal-mass particles decay,
\begin{align*}
  A &\rightarrow B + C\\
  A^\prime & \rightarrow B^\prime + C^\prime, 
\end{align*} 
and where one daughter from each parent, $B$ or $B^\prime$, is a
visible particle, and the other, $C$ or $C^\prime$ is not observed.
Since the $C$s are invisible their individual four-momenta are not
known. However the vector sum ${\bf p}_{\rm T}^\Sigma$ of the
transverse momentum components of $C$ and $C^\prime$ can be determined
from momentum conservation in the plane perpendicular to the beam.

For any given event \mttwo{} is defined to be the maximal possible
mass of the parent particle $A$ consistent with the constraints; that
is \mttwo{} provides the greatest lower bound on $m_A=m_{A^\prime}$
given the experimental observables~\cite{Cheng:2008hk}.

In the context of the di-Higgs decay \eqref{eq:hhbbtautau} the
dominant background process \eqref{eq:ttbarbg} satisfies the
assumptions under which \mttwo{} is useful: the dileptonic (di-tau)
\ttbar{} background involves the pair-production of identical-mass
parents; and each of which decays to a final state which contains
visible particles (the $b$ jets, and visible $\tau{}$ decay products)
and invisible particles (the neutrinos both from the $W$ decays and
from the leptonic or hadronic $\tau{}$ decays).  We can therefore
build a kinematical variable from the observed final state particles
which is bounded above by the top quark mass for the \ttbar{}
background, but remains unbounded above for the di-Higgs signal
process.

The \mttwo{} variable can be explicitly
constructed~\cite{Lester:1999tx} as
\begin{multline}
  \label{eq:mttwodef}
  \mttwo(m_B, m_{B'}, {\bf b}_{\rm T}, {\bf b}'_{\rm T}, {\bf p}_{\rm T}^\Sigma, m_{C}, m_{C'}) \\
  \equiv \min_{ {\bf c}_{\rm T} + {\bf c}'_{\rm T} = {\bf p}_{\rm
      T}^\Sigma } \left\{ \max { \left( \mt, \mt' \right) } \right\} ,
\end{multline}
where $\mt$ is the transverse mass constructed from $m_B$, ${m}_{C}$,
${\bf b}_{\rm T}$ and ${\bf c}_{\rm T}$, while $\mt'$ is the
transverse mass constructed from $m_{B'}$, ${m}_{{C}'}$, ${\bf
  b}'_{\rm T}$ and ${\bf c}'_{\rm T}$, and where the minimisation is
over all hypothesised transverse momenta ${\bf c}_{\rm T}$ and ${\bf
  c}'_{\rm T}$ for the invisible particles which sum to the constraint
${\bf p}_{\rm T}^\Sigma$, which is usually the observed missing
transverse momentum $\mptvec$.  The transverse mass \mt{} is itself
defined by
\begin{equation}\nonumber
%  \label{eq:mtdef}
  \mt^2({\bf b}_{\rm T}, {\bf c}_{\rm T}, m_b, m_c) 
  \equiv m_b^2 + m_{c}^2 + 2 \left( e_b e_c - {\bf b}_{\rm T} \cdot {\bf c}_{\rm T} \right) ,
\end{equation}
where the `transverse energy' $e$ for each particle is defined by
\begin{equation}\nonumber
%\label{eq:et_noz}
e^2 = {m^2 + {\bf p}^2_{\rm T}} .\
\end{equation}
Variants\footnote{See Ref.~\cite{Barr:2010zj} for a recent review, and
  Ref.~\cite{Barr:2011xt} for examples and categorisation.} of
\mttwo{} address cases where some or all of the $A$, $B$, $A^\prime$
or $B^\prime$ particles are composed of four-vector sums. Such
variants are designed for more complicated $n$-body decays with $n>2$
or for the case of sequential decays with on-shell intermediates.
While these mass-bounding variables were originally proposed to gain
sensitivity to the masses of new particles at hadron colliders, they
have also proved effective in
searches~\cite{Barr:2009wu,daCosta:2011qk,Aad:2011ib,Aad:2012pxa,Aad:2012uu}.

For the $hh\rightarrow b\bar{b}\tau^+ \tau^-$ case, an appropriate
variable is constructed as follows.  The $b$ jets resulting from each
of the two top quark decays enter \eqref{eq:mttwodef} as the visible
particles $B$ and $B^\prime$. The components $C$ and $C^\prime$ in
\eqref{eq:mttwodef} which form the transverse momentum constraint
should then be the sum of the decay products of the $W$ bosons.  The
appropriate vector sum ${\bf p}_{\rm T}^\Sigma$ for the constraint in
\eqref{eq:mttwodef} contains both visible and invisible components,
\begin{align}
  \label{eq:transverse_constraint}
  {\bf p}_{\rm T}^{\Sigma} &\equiv \mptvec + {\bf p}_{\rm T}^{\rm vis}(\tau) 
                                         + {\bf p}_{\rm T}^{\rm vis}(\tau') \\
                         &= {\bf p}_{\rm T}(W)
                          + {\bf p}_{\rm T}(W') , \nonumber
\end{align}
where the first line sums the missing transverse momentum \mptvec{}
(from all neutrinos from the leptonic $W$ decays, including subsequent
leptonic or hadronic $\tau$ decays), and the visible transverse
momentum from each of the two reconstructed $\tau$ candidates.

%%%%%%%%%%%%%%%%%%%%%%%%%%%%%%%%%%%%%%%%%%
%
%
\begin{table*}[!t]
\renewcommand\arraystretch{1.3}
\begin{tabular}{|| l | C{2cm} | C{2cm} | C{2cm} | C{2cm} | C{4cm} ||}
  \hline
                     & signal & \multicolumn{3}{c|}{backgrounds} & \\
\cline{3-5}
  cross section {[fb]} & $hh$   & $b\bar b W^+W^-$ &  $b\bar b \tau^+\tau^-$ & $b\bar b\tau^+\tau^-$ ew. & $S/B$ \\
  \hline
  Before cuts & 13.89  & 10792  & 2212 & 82.3 & $1.06\times 10^{-3}$ \\
  After trigger &  1.09 & 1966  & 372 & 15.0  & $0.463\times 10^{-3}$ \\
  After event selection & 0.248   & 383.0  & 43.7 & 2.08  & $0.578  \times 10^{-3}$ \\
  After $m(\tau^+\tau^-)$ cut & 0.164 [0.128]  & 107.7 [107.4] & 4.62  [16.0] & 0.316 [0.789]& $1.46 \times 10^{-3}$ [$ 1.02 \times 10^{-3}$]\\
  After $m(b\bar b)$ cut &  0.118 [0.093] &  28.7 [29.1] & 0.973 [4.03] & 0.062 [0.351] & $3.98\times 10^{-3}$ [$ 2.79 \times 10^{-3}$] \\ \hline
  After $\ptbbbar$ cut &  0.055 [0.041] &  0.475 [0.480] & 0.037 [0.247]
  & 0.013 [0.079] & $0.105$ [$ 0.050 $] \\
  After $\mttwo$ cut &  0.047 [0.034] &  0.147 [0.194] & 0.029 [0.204]
  & 0.012 [0.074] & $0.250$ [$ 0.072 $] \\
  \hline
\end{tabular}
\caption{\label{tab:bbtautau} Cross sections for $hh$ the signal and 
  for the dominant backgrounds after various selection criteria have 
  been applied. The $b\bar b W^+W^-$ column considers only the 
  decay of $W$ bosons to $\tau$ leptons,
  and already includes the corresponding branching ratios.
  The final column shows the signal to background ratio. The
  numbers in brackets follow from a more conservative $\tau\tau$ mass
  reconstruction, described in the text. The last rows correspond to
  exemplary
  cuts $p_{{\text{T}},b\bar b}\geq 175\GeV$ followed by $m_{\text{T2}}\geq 125\GeV$.
}
\end{table*}
%%%%%%%%%%%%%%%%%%%%%%%%%%%%%%%%%%%%%%%%%%

The resulting variable
\begin{equation}
  \label{eq:mttwobb} 
  \mttwo\left(m_b, m_b^\prime, {\bf b}_{\rm T}, {\bf b}'_{\rm T}, 
    {\bf p}_{\rm T}^\Sigma, m^{\rm vis}(\tau), m^{\rm vis}(\tau^\prime)\right)
\end{equation}
is by construction bounded above by $m_t$ for the \ttbar{} background
process (in the narrow width approximation, and in the absence of
detector resolution effects). By contrast, for the $hh$ signal the
\mttwo{} distribution can reach very large values, in principle up to
$\sqrt{s}/2$.

%%%%%%%%%%%%%%%%%%%%%%%%%%%%%%%%%%%%%%
\mysection{Elements of the Analysis}
%%%%%%%%%%%%%%%%%%%%%%%%%%%%%%%%%%%%%%
%%%%%%%%%%%%%%%%%%%%%%%%%%%%%%%%%%%%%%
\subsection{Detector simulation}
%%%%%%%%%%%%%%%%%%%%%%%%%%%%%%%%%%%%%%
We model the effects of detector resolution and efficiency using a
custom detector simulation based closely on the ATLAS `Krak\'ow'
parameterisation~\cite{ATL-PHYS-PUB-2013-004}.  The parameters
employed provide conservative estimates of the ATLAS detector
performance for the phase-II high-luminosity LHC machine (HL-LHC),
which is expected to deliver an integrated luminosity of $3\,\invab$
to each of the two general-purpose experiments.  In particular we
model pile-up (at $\mu=80$) and \sumEt{} dependent resolutions for
jets and for~$\ptmiss$.

Jets are reconstructed with the anti-$k_t$ jet clustering
algorithm~\cite{Cacciari:2008gp,Cacciari:2011ma} with radius parameter
$0.6$.  Tau lepton reconstruction efficiencies and fake rates are
included, based on Ref.~\cite{ATL-PHYS-PUB-2013-004}, as are jet
resolutions, and $b$-jet efficiencies and fake rates.

%%%%%%%%%%%%%%%%%%%%%%%%%%%%%%%%%%%%%%
\subsection{Event generation}
%%%%%%%%%%%%%%%%%%%%%%%%%%%%%%%%%%%%%%
To generate the signal and background events we closely follow
Ref.~\cite{Dolan:2012rv} (details of the comparison of the signal
Monte Carlo that underlies this study and comparisons against earlier
results can be found therein). Signal events $p(g)p(g) \to hh+X$
(which dominate the inclusive $hh$ cross section) are generated with a
combination of the {\sc{Vbfnlo}} \cite{Arnold:2008rz} and
{\sc{FeynArts}}/{\sc{FormCalc}}/{\sc{LoopTools}}
\cite{Hahn:2000kx,Hahn:1998yk} frameworks. We generate events in the
Les Houches standard~\cite{Boos:2001cv} which we pass to
{\sc{Herwig++}}~\cite{Bahr:2008pv} for showering and hadronisation of
the selected $h\to b\bar b,\tau^+\tau^-$ final states. We use a flat
NLO QCD factor to account for higher order perturbative corrections by
effectively normalizing to an inclusive cross section of $\sigma =
33.89$~fb~\cite{Baglio:2012np,Dawson:1998py}.

The QCD and electroweak $b\bar b \tau^+\tau^-$ backgrounds are
generated with {\sc{Sherpa}}~\cite{Gleisberg:2008ta} and the $t\bar t$
background of Eq.~\eqref{eq:ttbarbg} is generated with
{\sc{MadEvent}}~5~\cite{Alwall:2011uj}. The $b\bar b W^+W^-$ NLO cross
sections have been computed in Ref.~\cite{Denner:2010jp} (we use
$K\simeq 1.5$ and specify $W\to \tau\nu_\tau$ in {\sc{Herwig++}}
during showering and hadronisation to increase the efficiency for the
cut selection), for the mixed QCD/electroweak and the purely
electroweak contributions we use the corrections to $Zb\bar b$
($K\simeq 1.4$) and $ZZ$ ($K\simeq 1.6$) production using
{\sc{Mcfm}}~\cite{zbb1,zbb2,Campbell:2011bn}.

%%%%%%%%%%%%%%%%%%%%%%%%%%%%%%%%%%%%%%%%%%%%%%
\begin{figure*}
 \begin{center}
  \subfigure[]{\includegraphics[width=0.4\linewidth]{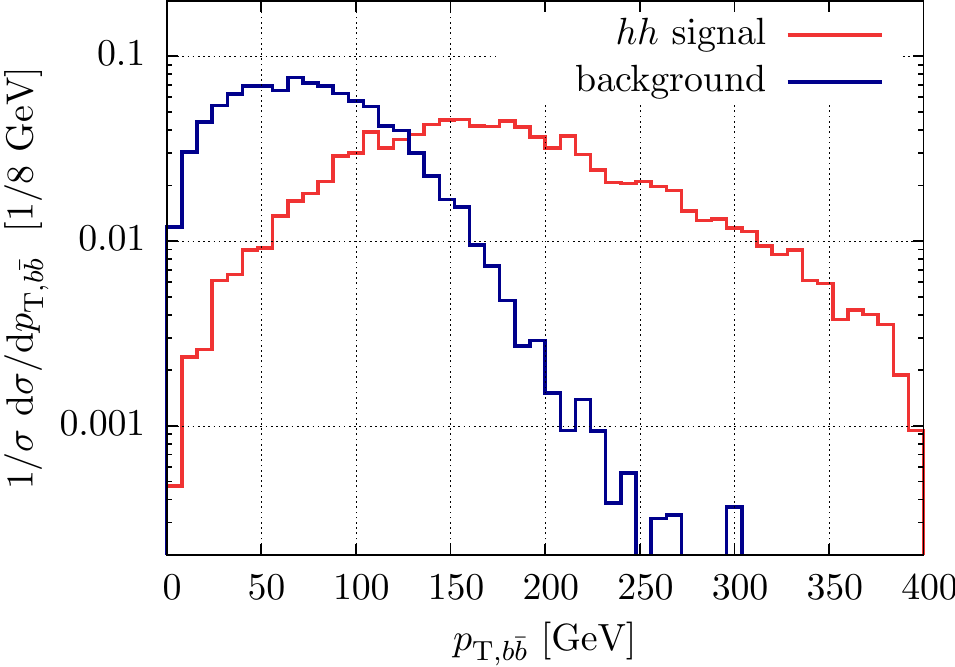}
    \label{fig:ptbb}}
  \hspace{1cm}
  \subfigure[]{\includegraphics[width=0.4\linewidth]{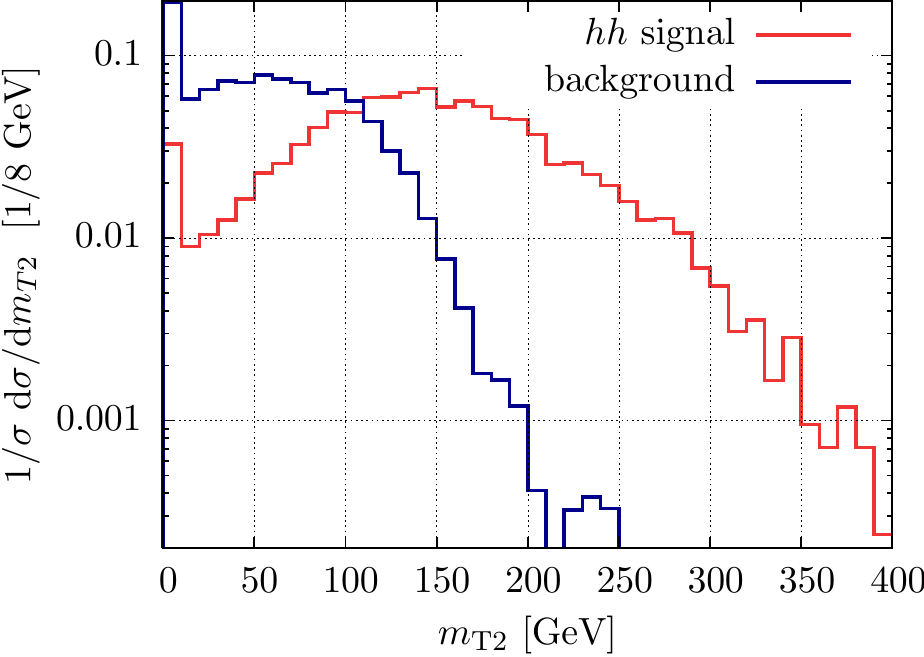}
    \label{fig:mt2}}
  \caption{Transverse momentum distribution of the reconstructed Higgs
    (i.e. the $b\bar b$ pair) and the $m_{\text{T2}}$ distribution
    after the analysis steps described in the text have been carried
    out (see also Tab.~\ref{tab:bbtautau}) but before cuts on either
    $m_{\text{T2}}$ or $p_{\text{T},b\bar b}$ have been applied.}
\label{fig:mt2ptbb}
\end{center}
\end{figure*}
%%%%%%%%%%%%%%%%%%%%%%%%%%%%%%%%%%%%%%%%%%%%%%
%%%%%%%%%%%%%%%%%%%%%%%%%%%%%%%%%%%%%%
\mysubsection{Event selection}
\label{sec:cuts} 
%%%%%%%%%%%%%%%%%%%%%%%%%%%%%%%%%%%%%%
Events are assumed to pass the trigger if there are at least two
$\tau$s with visible $\pt>40\GeV$ or at least one $\tau$ with visible
$\pt>60\GeV$.  Both leptonic and hadronic decays of $\tau$s are
included.  Selected events are required to have exactly two
reconstructed $\tau$s (leptonic or hadronic) and exactly two
reconstructed and $b$-tagged jets.

The reconstruction of the di-tau mass is important in discriminating
the $h\to \tau^+\tau^-$ from the $Z\to \tau^+\tau^-$ background.  The
LHC experiments typically employ sophisticated mass-reconstruction
methods which include kinematic constraints but also likelihood
functions or multi-variate techniques trained to mitigate against
detector resolution~\cite{Chatrchyan:2012vp,Aad:2012mea}.  We use a
simpler, purely kinematic reconstruction of the di-tau mass, which is
not expected to perform as well as the techniques used by the
experiments in the presence of detector smearing.  To estimate the
systematic impact of the $\tau$ reconstruction on $h\to \tau\tau$
selection, we perform the same \mboundgeneral{} reconstruction with
and without simulation of the \ptmiss{} resolution.  The more
sophisticated techniques used by the experiments which mitigate
against detector resolution can be expected to lie between our two
estimates.

In each case we construct a $\tau^+\tau^-$ invariant mass bound
$\mboundgeneral$ using the greatest lower bound \mboundfull{} on $m_h$
given the visible momenta, \ptmiss{} and $m_\tau$
constraints~\cite{Barr:2011he}. When detector smearing leads to events
where \mboundfull{} does not exist, the $\tau$ mass constraints are
dropped, and the resulting transverse mass \mt{} is used as the
greatest lower bound $\mboundgeneral$ on $m_h$.

In each case we require that \mboundgeneral{} lie within a 50\,GeV
window.  In the analysis without \ptmiss{} smearing we choose $100
\GeV< m_{\tau\tau} < 150 \GeV$, while we select $80 \GeV< m_{\tau\tau}
< 130 \GeV$ when smearing is included. Note that in the latter case
$Z\to\tau^+ \tau^-$ is a large contamination of the signal region
defined by the invariant mass windows. By calibrating the Higgs mass
reconstruction from $h\to \tau^+\tau^-$ as already presently performed
in the $Z\to\tau^+\tau^-$ case~\cite{taurec1,taurec2}, this
contamination could be reduced.

The $b\bar b$ invariant mass is calculated from the four-vector sum of
the two $b$-tagged jets.  Events are selected if they satisfy $100
\GeV < m_{bb}< 150\GeV$.

%%%%%%%%%%%%%%%%%%%%%%%%%%%%%%%%%%%%%%
\mysection{Results}
%%%%%%%%%%%%%%%%%%%%%%%%%%%%%%%%%%%%%%
The numbers of events passing each of the selection criteria are
tabulated in Tab.~\ref{tab:bbtautau}.  We find that the transverse
momentum and $m_{\text{T2}}$ observables are necessary for background
suppression, and, hence, for a potentially successful measurement of
the di-Higgs final state in a hadronically busy environment. The
normalized $\mttwo$ and $\ptbbbar$ distributions after the selection
shown in Tab.~\ref{tab:bbtautau} are plotted in
Fig.~\ref{fig:mt2ptbb}.  It can be seen that each of the two variables
offers good signal versus background discrimination at the large
integrated luminosities anticipated at the high luminosity LHC.  We
also observe that, $\mttwo$ and $\ptbbbar$ encode orthogonal
information and they can be combined towards an optimised search
strategy.

We find it is straightforward to obtain signal-to-background $(S/B)$
ratios of $\sim 1/5$ while retaining acceptably large signal cross
section.  These ratios are re-expressed in Fig.~\ref{fig:mt2ptbb}
which depicts the luminosity contours that are necessary to claim a
$5\sigma$ discovery of di-Higgs production on the basis of a simple
`cut and count' experiment that makes the rectangular cut requirements
that both $\ptbbbar > \ptbbbar({\rm cut)}$ and $\mttwo>\mttwo({\rm
  cut})$. Both axes stop at rather low values of
$\left(\ptbbbar,\,\mttwo\right)$ since a tighter selection would be
dependent on the tail of the $t\bar{t}$ distribution where
$S/\sqrt{B}$ does not provide an appropriate indicator of
sensitivity. We find that the HL-LHC has good sensitivity to the $hh$
production at high luminosity.  For an example selection we obtain a
cross section measurement in the 30\% range (including the statistical
background uncertainty).

The sensitivity to the Higgs trilinear coupling follows from
destructive interference with other SM diagrams (see
Ref.~\cite{Dolan:2012rv}), such that
\begin{equation}
  \label{eq:triqual}
  \lambda \gtrless\lambda_{\text{SM}} \; \Longrightarrow \; 
  \sigma(hh) \lessgtr
  \sigma(hh)_{\text{SM}} \,.
\end{equation} 
Using the full parton-level $p(g)p(g)\to hh+X$
calculation~\cite{Dolan:2012rv} we find that the quoted 30\% cross
section uncertainty translates into 60\% level sensitivity to the
Higgs trilinear coupling in the part of the $p_{{\text{T}},b \bar b}$
distribution which is relevant for this analysis, $p_{{\text{T}},b\bar
  b}\gtrsim 180\GeV$.

%%%%%%%%%%%%%%%%%%%%%%%%%%%%%%%%%%%%%%%%%%%%%%
\begin{figure}[!t]
  \parbox{0.48\textwidth}{
    \vskip-0.5cm
    \includegraphics[width=0.48\textwidth]{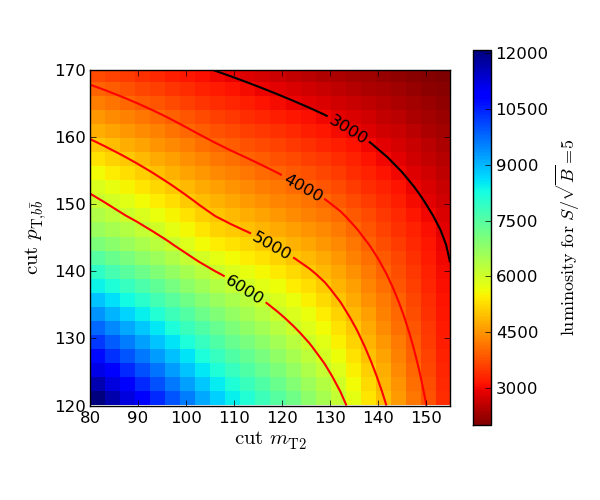}
  }
  \vskip-0.5cm
  \caption{Luminosity in \invfb{} required to reach $S/\sqrt{B}=5$ for
    di-Higgs production based on simple rectangular cuts on $\ptbbbar$
    and $\mttwo$.  Numbers in red show luminosities that would require
    a combination of the ATLAS and CMS data sets from a $3\,\invab$
    high luminosity LHC.}
  \label{fig:mt2ptbbcut}
\end{figure}
%%%%%%%%%%%%%%%%%%%%%%%%%%%%%%%%%%%%%%%%%%%%%%

As an alternative to a `cut and count' analysis we construct a two
dimensional likelihood from $(\mttwo,\ptbbbar)$ to obtain an estimate
of the maximal sensitivity that is encoded in these observables,
including their correlation~\cite{edwards}.  Figures~\ref{fig:mt2ptbb}
and \ref{fig:mt2ptbbcut} show that the best sensitivity will result
from energetic events either with large \mttwo{} or large \ptbbbar{}
or both.  Using the likelihood method we find fractional uncertainty
in the cross section of
\begin{multline}
  \label{eq:binned}
  \left[\sigma/\sigma(hh)_{\text{SM}}\right]_{\text{excl}}\simeq
  0.37~[1.00]\\
  \hbox{for $3\,\invab$ at 95\% confidence level}
\end{multline}
using the CL(s) method~\cite{Read:2002hq}.  The sensitivity to the
$pp\to hh+X$ cross section as captured in Eq.~\eqref{eq:binned} can be
rephrased into an expected upper 95\% CL bound on the Higgs
self-interaction in the $b\bar{b} \tau^+ \tau^-$ channel via
Eq.~\eqref{eq:triqual}.  For a background-only hypothesis with no true
$hh$ production we would find a limit on the self-coupling of
\begin{equation}\nonumber
  \lambda> \lambda\big|_{95\%~\text{CL}}^{3000/{\text{fb}}} \simeq  
  3.0~[1.0]\times \lambda_{\text{SM}}\,,
\end{equation}
where it should be noted that a 95\% CL of $3\,\lambda_{\text{SM}}$ is
more stringent than the case of $1\,\lambda_{\text{SM}}$ due to the
destructive interference~\eqref{eq:triqual}.

A measurement of $\lambda$ at this level would be sufficient constrain
a wide range of scenarios of electroweak symmetry breaking, such as
composite-Higgs models and pseudo-dilaton models which can lead to
large increases in the Higgs self-coupling.  While the limit might be
somewhat degraded by additional systematic uncertainties in background
determination, it also has the potential to be improved by using a
subjet analysis~\cite{Dolan:2012rv}, and/or by using the more
sophisticated di-tau mass reconstruction techniques already employed
by the LHC experiments.

%%%%%%%%%%%%%%%%%%%%%%%%%%%%%%%%%%%%%%
\mysection{Summary and Conclusions}
%%%%%%%%%%%%%%%%%%%%%%%%%%%%%%%%%%%%%%
Following the discovery of a Higgs boson, one of the top priorities at
the LHC is to address the mechanism of electroweak symmetry breaking
at a more fundamental level.

In this work we have shown that the $3\,\invab{}$ high luminosity LHC
will have sensitivity to the Higgs self-coupling in the favoured
$b\bar b\tau^+\tau^-$ channel using two simple kinematic variables
$\mttwo$ and $\ptbbbar$ each of which independently suppresses the
dominant $t\bar t$ background.

We have used parameterised detector simulations of the ATLAS detector
as expected for a high-luminosity environment throughout.

Using a two dimensional log-likelihood approach, the null hypothesis
of $\sigma(hh)=0$ would constrain the Higgs trilinear coupling to
$\lambda \gtrsim 3.0~[1.0]~\lambda_{\text{SM}}$ at the 95\% confidence
level.  An exemplary cross section measurement with 30\% precision
translates into a measurement of $\lambda$ at the 60\% level.

However further improvements to the presented analysis are possible:
\begin{enumerate}
\item Jet substructure techniques allow one to narrow the invariant
  $b\bar b$ mass window \cite{Dolan:2012rv}, thus leading to a larger
  rejection of the $b\bar b W^+W^-$ and $b\bar b\tau^+ \tau^-$
  backgrounds.
\item Calibrated $h\to \tau^+\tau^- $ taggers~\cite{taurec1,taurec2}
  will highly suppress the $b\bar b \tau^+\tau^-$ backgrounds and also
  further reduce the $t\bar t$ background.
\item We have used a definition of $m_{\text{T2}}$ which does not
  include any information about the $\tau$ lepton momenta other than
  the sum of their $\pt$ in \eqref{eq:transverse_constraint}.  Using
  calibrated taggers, further kinematic information is available by
  modifying Eq.~\eqref{eq:mttwobb} through pairing $\tau$ and $b$
  objects, and exploiting the Jacobian peak of $m_{b\tau}$ in the top
  decay~\cite{Plehn:2011tf}. One can pair the hardest $\tau$ with that
  particular $b$ jet that yields an $m_{b\tau}$ value that is closer
  to $140\GeV$ (the maximum of the Jacobian peak). The leftover $b$
  jet is paired with the softer $\tau$, and the following
  substitutions used in Eq.~\eqref{eq:mttwobb}
  \begin{align*}
    m_b&\to m_{b\tau}\,,\\
    {\bf{b}}^{'}&\to {\bf{b}}^{'}+{\bf{\tau}}^{'}\,,\\
    {\bf p}_{\rm T}^{\Sigma} & \to {\slashed{p}_T} \,,\\
    m^{\text{vis}}&\to m^{\text{vis}}_{b\tau}\,,
  \end{align*}
  where the latter line indicates that the visible $\tau$ decay
  products are included in the invariant visible mass definition.
\end{enumerate}

A combination of such techniques can be used by the LHC experiments to
gain improved sensitivity to the Higgs self-coupling --- and hence to
the nature of electroweak symmetry breaking.

\bigskip

%%%%%%%%%%%%%%%%%%%%%%%%%%%%%%%%%%%%%%
\begin{acknowledgments}
  {\it Acknowledgments.} This work was supported by the Science and
  Technology Research Council of the United Kingdom, by Merton
  College, Oxford, and by the Institute for Particle Physics
  Phenomenology Associate Scheme. AJB thanks NORDITA and the IPPP for
  their hospitality during the preparation of this paper. CE thanks
  the IPPP for hospitality during the time when this work was
  completed. We thank Margarete M\"uhlleitner and Michael Spira for
  interesting discussions during the 2013 Les Houches workshop. Also,
  we thank Mike Johnson, Peter Richardson, and Ewan Steele for
  computing support and their patience.
\end{acknowledgments}
%%%%%%%%%%%%%%%%%%%%%%%%%%%%%%%%%%%%%%

\bibliography{hhh_mt2}

\begin{thebibliography}{60}
\expandafter\ifx\csname natexlab\endcsname\relax\def\natexlab#1{#1}\fi
\expandafter\ifx\csname bibnamefont\endcsname\relax
  \def\bibnamefont#1{#1}\fi
\expandafter\ifx\csname bibfnamefont\endcsname\relax
  \def\bibfnamefont#1{#1}\fi
\expandafter\ifx\csname citenamefont\endcsname\relax
  \def\citenamefont#1{#1}\fi
\expandafter\ifx\csname url\endcsname\relax
  \def\url#1{\texttt{#1}}\fi
\expandafter\ifx\csname urlprefix\endcsname\relax\def\urlprefix{URL }\fi
\providecommand{\bibinfo}[2]{#2}
\providecommand{\eprint}[2][]{\url{#2}}

\bibitem[{\citenamefont{Aad et~al.}(2012{\natexlab{a}})}]{Aad:2012tfa}
\bibinfo{author}{\bibfnamefont{G.}~\bibnamefont{Aad}} \bibnamefont{et~al.}
  (\bibinfo{collaboration}{ATLAS Collaboration}), \bibinfo{journal}{Phys.Lett.}
  \textbf{\bibinfo{volume}{B716}}, \bibinfo{pages}{1}
  (\bibinfo{year}{2012}{\natexlab{a}}), \eprint{1207.7214}.

\bibitem[{\citenamefont{Chatrchyan
  et~al.}(2012{\natexlab{a}})}]{Chatrchyan:2012ufa}
\bibinfo{author}{\bibfnamefont{S.}~\bibnamefont{Chatrchyan}}
  \bibnamefont{et~al.} (\bibinfo{collaboration}{CMS Collaboration}),
  \bibinfo{journal}{Phys.Lett.} \textbf{\bibinfo{volume}{B716}},
  \bibinfo{pages}{30} (\bibinfo{year}{2012}{\natexlab{a}}), \eprint{1207.7235}.

\bibitem[{\citenamefont{Higgs}(1964{\natexlab{a}})}]{Higgs:1964ia}
\bibinfo{author}{\bibfnamefont{P.~W.} \bibnamefont{Higgs}},
  \bibinfo{journal}{Phys.Lett.} \textbf{\bibinfo{volume}{12}},
  \bibinfo{pages}{132} (\bibinfo{year}{1964}{\natexlab{a}}).

\bibitem[{\citenamefont{Higgs}(1964{\natexlab{b}})}]{Higgs:1964pj}
\bibinfo{author}{\bibfnamefont{P.~W.} \bibnamefont{Higgs}},
  \bibinfo{journal}{Phys.Rev.Lett.} \textbf{\bibinfo{volume}{13}},
  \bibinfo{pages}{508} (\bibinfo{year}{1964}{\natexlab{b}}).

\bibitem[{\citenamefont{Guralnik et~al.}(1964)\citenamefont{Guralnik, Hagen,
  and Kibble}}]{Guralnik:1964eu}
\bibinfo{author}{\bibfnamefont{G.}~\bibnamefont{Guralnik}},
  \bibinfo{author}{\bibfnamefont{C.}~\bibnamefont{Hagen}}, \bibnamefont{and}
  \bibinfo{author}{\bibfnamefont{T.}~\bibnamefont{Kibble}},
  \bibinfo{journal}{Phys.Rev.Lett.} \textbf{\bibinfo{volume}{13}},
  \bibinfo{pages}{585} (\bibinfo{year}{1964}).

\bibitem[{\citenamefont{Englert and Brout}(1964)}]{Englert:1964et}
\bibinfo{author}{\bibfnamefont{F.}~\bibnamefont{Englert}} \bibnamefont{and}
  \bibinfo{author}{\bibfnamefont{R.}~\bibnamefont{Brout}},
  \bibinfo{journal}{Phys.Rev.Lett.} \textbf{\bibinfo{volume}{13}},
  \bibinfo{pages}{321} (\bibinfo{year}{1964}).

\bibitem[{\citenamefont{Coleman and Weinberg}(1973)}]{Coleman:1973jx}
\bibinfo{author}{\bibfnamefont{S.~R.} \bibnamefont{Coleman}} \bibnamefont{and}
  \bibinfo{author}{\bibfnamefont{E.~J.} \bibnamefont{Weinberg}},
  \bibinfo{journal}{Phys.Rev.} \textbf{\bibinfo{volume}{D7}},
  \bibinfo{pages}{1888} (\bibinfo{year}{1973}).

\bibitem[{\citenamefont{Plehn and Rauch}(2005)}]{Plehn:2005nk}
\bibinfo{author}{\bibfnamefont{T.}~\bibnamefont{Plehn}} \bibnamefont{and}
  \bibinfo{author}{\bibfnamefont{M.}~\bibnamefont{Rauch}},
  \bibinfo{journal}{Phys.Rev.} \textbf{\bibinfo{volume}{D72}},
  \bibinfo{pages}{053008} (\bibinfo{year}{2005}), \eprint{hep-ph/0507321}.

\bibitem[{\citenamefont{Glover and van~der Bij}(1988)}]{Glover:1987nx}
\bibinfo{author}{\bibfnamefont{E.~N.} \bibnamefont{Glover}} \bibnamefont{and}
  \bibinfo{author}{\bibfnamefont{J.}~\bibnamefont{van~der Bij}},
  \bibinfo{journal}{Nucl.Phys.} \textbf{\bibinfo{volume}{B309}},
  \bibinfo{pages}{282} (\bibinfo{year}{1988}).

\bibitem[{\citenamefont{Dicus et~al.}(1988)\citenamefont{Dicus, Kao, and
  Willenbrock}}]{Dicus:1987ic}
\bibinfo{author}{\bibfnamefont{D.~A.} \bibnamefont{Dicus}},
  \bibinfo{author}{\bibfnamefont{C.}~\bibnamefont{Kao}}, \bibnamefont{and}
  \bibinfo{author}{\bibfnamefont{S.~S.} \bibnamefont{Willenbrock}},
  \bibinfo{journal}{Phys.Lett.} \textbf{\bibinfo{volume}{B203}},
  \bibinfo{pages}{457} (\bibinfo{year}{1988}).

\bibitem[{\citenamefont{Plehn et~al.}(1996)\citenamefont{Plehn, Spira, and
  Zerwas}}]{Plehn:1996wb}
\bibinfo{author}{\bibfnamefont{T.}~\bibnamefont{Plehn}},
  \bibinfo{author}{\bibfnamefont{M.}~\bibnamefont{Spira}}, \bibnamefont{and}
  \bibinfo{author}{\bibfnamefont{P.}~\bibnamefont{Zerwas}},
  \bibinfo{journal}{Nucl.Phys.} \textbf{\bibinfo{volume}{B479}},
  \bibinfo{pages}{46} (\bibinfo{year}{1996}), \eprint{hep-ph/9603205}.

\bibitem[{\citenamefont{Djouadi et~al.}(1999)\citenamefont{Djouadi, Kilian,
  Muhlleitner, and Zerwas}}]{Djouadi:1999rca}
\bibinfo{author}{\bibfnamefont{A.}~\bibnamefont{Djouadi}},
  \bibinfo{author}{\bibfnamefont{W.}~\bibnamefont{Kilian}},
  \bibinfo{author}{\bibfnamefont{M.}~\bibnamefont{Muhlleitner}},
  \bibnamefont{and} \bibinfo{author}{\bibfnamefont{P.}~\bibnamefont{Zerwas}},
  \bibinfo{journal}{Eur.Phys.J.} \textbf{\bibinfo{volume}{C10}},
  \bibinfo{pages}{45} (\bibinfo{year}{1999}), \eprint{hep-ph/9904287}.

\bibitem[{\citenamefont{de~Florian and Mazzitelli}(2013)}]{deFlorian:2013uza}
\bibinfo{author}{\bibfnamefont{D.}~\bibnamefont{de~Florian}} \bibnamefont{and}
  \bibinfo{author}{\bibfnamefont{J.}~\bibnamefont{Mazzitelli}},
  \bibinfo{journal}{Phys.Lett.} \textbf{\bibinfo{volume}{B724}},
  \bibinfo{pages}{306} (\bibinfo{year}{2013}), \eprint{1305.5206}.

\bibitem[{\citenamefont{Grigo et~al.}(2013{\natexlab{a}})\citenamefont{Grigo,
  Hoff, Melnikov, and Steinhauser}}]{Grigo}
\bibinfo{author}{\bibfnamefont{J.}~\bibnamefont{Grigo}},
  \bibinfo{author}{\bibfnamefont{J.}~\bibnamefont{Hoff}},
  \bibinfo{author}{\bibfnamefont{K.}~\bibnamefont{Melnikov}}, \bibnamefont{and}
  \bibinfo{author}{\bibfnamefont{M.}~\bibnamefont{Steinhauser}},
  \bibinfo{journal}{Nucl.Phys.} \textbf{\bibinfo{volume}{B875}},
  \bibinfo{pages}{1} (\bibinfo{year}{2013}{\natexlab{a}}), \eprint{1305.7340}.

\bibitem[{\citenamefont{Baur et~al.}(2003)\citenamefont{Baur, Plehn, and
  Rainwater}}]{Baur:2002qd}
\bibinfo{author}{\bibfnamefont{U.}~\bibnamefont{Baur}},
  \bibinfo{author}{\bibfnamefont{T.}~\bibnamefont{Plehn}}, \bibnamefont{and}
  \bibinfo{author}{\bibfnamefont{D.~L.} \bibnamefont{Rainwater}},
  \bibinfo{journal}{Phys.Rev.} \textbf{\bibinfo{volume}{D67}},
  \bibinfo{pages}{033003} (\bibinfo{year}{2003}), \eprint{hep-ph/0211224}.

\bibitem[{\citenamefont{Dolan et~al.}(2012)\citenamefont{Dolan, Englert, and
  Spannowsky}}]{Dolan:2012rv}
\bibinfo{author}{\bibfnamefont{M.~J.} \bibnamefont{Dolan}},
  \bibinfo{author}{\bibfnamefont{C.}~\bibnamefont{Englert}}, \bibnamefont{and}
  \bibinfo{author}{\bibfnamefont{M.}~\bibnamefont{Spannowsky}},
  \bibinfo{journal}{JHEP} \textbf{\bibinfo{volume}{1210}}, \bibinfo{pages}{112}
  (\bibinfo{year}{2012}), \eprint{1206.5001}.

\bibitem[{\citenamefont{Papaefstathiou
  et~al.}(2013)\citenamefont{Papaefstathiou, Yang, and
  Zurita}}]{Papaefstathiou:2012qe}
\bibinfo{author}{\bibfnamefont{A.}~\bibnamefont{Papaefstathiou}},
  \bibinfo{author}{\bibfnamefont{L.~L.} \bibnamefont{Yang}}, \bibnamefont{and}
  \bibinfo{author}{\bibfnamefont{J.}~\bibnamefont{Zurita}},
  \bibinfo{journal}{Phys.Rev.} \textbf{\bibinfo{volume}{D87}},
  \bibinfo{pages}{011301} (\bibinfo{year}{2013}), \eprint{1209.1489}.

\bibitem[{\citenamefont{Dolan et~al.}(2013)\citenamefont{Dolan, Englert, and
  Spannowsky}}]{Dolan:2012ac}
\bibinfo{author}{\bibfnamefont{M.~J.} \bibnamefont{Dolan}},
  \bibinfo{author}{\bibfnamefont{C.}~\bibnamefont{Englert}}, \bibnamefont{and}
  \bibinfo{author}{\bibfnamefont{M.}~\bibnamefont{Spannowsky}},
  \bibinfo{journal}{Phys.Rev.} \textbf{\bibinfo{volume}{D87}},
  \bibinfo{pages}{055002} (\bibinfo{year}{2013}), \eprint{1210.8166}.

\bibitem[{\citenamefont{Baglio et~al.}(2013)\citenamefont{Baglio, Djouadi,
  Grober, Muhlleitner, Quevillon et~al.}}]{Baglio:2012np}
\bibinfo{author}{\bibfnamefont{J.}~\bibnamefont{Baglio}},
  \bibinfo{author}{\bibfnamefont{A.}~\bibnamefont{Djouadi}},
  \bibinfo{author}{\bibfnamefont{R.}~\bibnamefont{Grober}},
  \bibinfo{author}{\bibfnamefont{M.}~\bibnamefont{Muhlleitner}},
  \bibinfo{author}{\bibfnamefont{J.}~\bibnamefont{Quevillon}},
  \bibnamefont{et~al.}, \bibinfo{journal}{JHEP}
  \textbf{\bibinfo{volume}{1304}}, \bibinfo{pages}{151} (\bibinfo{year}{2013}),
  \eprint{1212.5581}.

\bibitem[{\citenamefont{Goertz et~al.}(2013)\citenamefont{Goertz,
  Papaefstathiou, Yang, and Zurita}}]{Goertz:2013kp}
\bibinfo{author}{\bibfnamefont{F.}~\bibnamefont{Goertz}},
  \bibinfo{author}{\bibfnamefont{A.}~\bibnamefont{Papaefstathiou}},
  \bibinfo{author}{\bibfnamefont{L.~L.} \bibnamefont{Yang}}, \bibnamefont{and}
  \bibinfo{author}{\bibfnamefont{J.}~\bibnamefont{Zurita}},
  \bibinfo{journal}{JHEP} \textbf{\bibinfo{volume}{1306}}, \bibinfo{pages}{016}
  (\bibinfo{year}{2013}), \eprint{1301.3492}.

\bibitem[{\citenamefont{Cao et~al.}(2013)\citenamefont{Cao, Heng, Shang, Wan,
  and Yang}}]{Cao:2013si}
\bibinfo{author}{\bibfnamefont{J.}~\bibnamefont{Cao}},
  \bibinfo{author}{\bibfnamefont{Z.}~\bibnamefont{Heng}},
  \bibinfo{author}{\bibfnamefont{L.}~\bibnamefont{Shang}},
  \bibinfo{author}{\bibfnamefont{P.}~\bibnamefont{Wan}}, \bibnamefont{and}
  \bibinfo{author}{\bibfnamefont{J.~M.} \bibnamefont{Yang}},
  \bibinfo{journal}{JHEP} \textbf{\bibinfo{volume}{1304}}, \bibinfo{pages}{134}
  (\bibinfo{year}{2013}), \eprint{1301.6437}.

\bibitem[{\citenamefont{Gupta et~al.}(2013)\citenamefont{Gupta, Rzehak, and
  Wells}}]{Gupta:2013zza}
\bibinfo{author}{\bibfnamefont{R.~S.} \bibnamefont{Gupta}},
  \bibinfo{author}{\bibfnamefont{H.}~\bibnamefont{Rzehak}}, \bibnamefont{and}
  \bibinfo{author}{\bibfnamefont{J.~D.} \bibnamefont{Wells}}
  (\bibinfo{year}{2013}), \eprint{1305.6397}.

\bibitem[{\citenamefont{Grigo et~al.}(2013{\natexlab{b}})\citenamefont{Grigo,
  Hoff, Melnikov, and Steinhauser}}]{Grigo:2013rya}
\bibinfo{author}{\bibfnamefont{J.}~\bibnamefont{Grigo}},
  \bibinfo{author}{\bibfnamefont{J.}~\bibnamefont{Hoff}},
  \bibinfo{author}{\bibfnamefont{K.}~\bibnamefont{Melnikov}}, \bibnamefont{and}
  \bibinfo{author}{\bibfnamefont{M.}~\bibnamefont{Steinhauser}},
  \bibinfo{journal}{Nucl.Phys.} \textbf{\bibinfo{volume}{B875}},
  \bibinfo{pages}{1} (\bibinfo{year}{2013}{\natexlab{b}}), \eprint{1305.7340}.

\bibitem[{\citenamefont{Nhung et~al.}(2013)\citenamefont{Nhung, Muhlleitner,
  Streicher, and Walz}}]{Nhung:2013lpa}
\bibinfo{author}{\bibfnamefont{D.~T.} \bibnamefont{Nhung}},
  \bibinfo{author}{\bibfnamefont{M.}~\bibnamefont{Muhlleitner}},
  \bibinfo{author}{\bibfnamefont{J.}~\bibnamefont{Streicher}},
  \bibnamefont{and} \bibinfo{author}{\bibfnamefont{K.}~\bibnamefont{Walz}}
  (\bibinfo{year}{2013}), \eprint{1306.3926}.

\bibitem[{\citenamefont{Ellwanger}(2013)}]{Ellwanger:2013ova}
\bibinfo{author}{\bibfnamefont{U.}~\bibnamefont{Ellwanger}},
  \bibinfo{journal}{JHEP} \textbf{\bibinfo{volume}{1308}}, \bibinfo{pages}{077}
  (\bibinfo{year}{2013}), \eprint{1306.5541}.

\bibitem[{\citenamefont{Butterworth et~al.}(2008)\citenamefont{Butterworth,
  Davison, Rubin, and Salam}}]{Butterworth:2008iy}
\bibinfo{author}{\bibfnamefont{J.~M.} \bibnamefont{Butterworth}},
  \bibinfo{author}{\bibfnamefont{A.~R.} \bibnamefont{Davison}},
  \bibinfo{author}{\bibfnamefont{M.}~\bibnamefont{Rubin}}, \bibnamefont{and}
  \bibinfo{author}{\bibfnamefont{G.~P.} \bibnamefont{Salam}},
  \bibinfo{journal}{Phys.Rev.Lett.} \textbf{\bibinfo{volume}{100}},
  \bibinfo{pages}{242001} (\bibinfo{year}{2008}), \eprint{0802.2470}.

\bibitem[{\citenamefont{Plehn et~al.}(2010)\citenamefont{Plehn, Salam, and
  Spannowsky}}]{Plehn:2009rk}
\bibinfo{author}{\bibfnamefont{T.}~\bibnamefont{Plehn}},
  \bibinfo{author}{\bibfnamefont{G.~P.} \bibnamefont{Salam}}, \bibnamefont{and}
  \bibinfo{author}{\bibfnamefont{M.}~\bibnamefont{Spannowsky}},
  \bibinfo{journal}{Phys.Rev.Lett.} \textbf{\bibinfo{volume}{104}},
  \bibinfo{pages}{111801} (\bibinfo{year}{2010}), \eprint{0910.5472}.

\bibitem[{\citenamefont{Lester and Summers}(1999)}]{Lester:1999tx}
\bibinfo{author}{\bibfnamefont{C.}~\bibnamefont{Lester}} \bibnamefont{and}
  \bibinfo{author}{\bibfnamefont{D.}~\bibnamefont{Summers}},
  \bibinfo{journal}{Phys.Lett.} \textbf{\bibinfo{volume}{B463}},
  \bibinfo{pages}{99} (\bibinfo{year}{1999}), \eprint{hep-ph/9906349}.

\bibitem[{\citenamefont{Barr et~al.}(2003)\citenamefont{Barr, Lester, and
  Stephens}}]{Barr:2003rg}
\bibinfo{author}{\bibfnamefont{A.}~\bibnamefont{Barr}},
  \bibinfo{author}{\bibfnamefont{C.}~\bibnamefont{Lester}}, \bibnamefont{and}
  \bibinfo{author}{\bibfnamefont{P.}~\bibnamefont{Stephens}},
  \bibinfo{journal}{J.Phys.} \textbf{\bibinfo{volume}{G29}},
  \bibinfo{pages}{2343} (\bibinfo{year}{2003}), \eprint{hep-ph/0304226}.

\bibitem[{\citenamefont{Cheng and Han}(2008)}]{Cheng:2008hk}
\bibinfo{author}{\bibfnamefont{H.-C.} \bibnamefont{Cheng}} \bibnamefont{and}
  \bibinfo{author}{\bibfnamefont{Z.}~\bibnamefont{Han}},
  \bibinfo{journal}{JHEP} \textbf{\bibinfo{volume}{0812}}, \bibinfo{pages}{063}
  (\bibinfo{year}{2008}), \eprint{0810.5178}.

\bibitem[{\citenamefont{Barr and Lester}(2010)}]{Barr:2010zj}
\bibinfo{author}{\bibfnamefont{A.~J.} \bibnamefont{Barr}} \bibnamefont{and}
  \bibinfo{author}{\bibfnamefont{C.~G.} \bibnamefont{Lester}},
  \bibinfo{journal}{J.Phys.} \textbf{\bibinfo{volume}{G37}},
  \bibinfo{pages}{123001} (\bibinfo{year}{2010}), \eprint{1004.2732}.

\bibitem[{\citenamefont{Barr et~al.}(2011{\natexlab{a}})\citenamefont{Barr,
  Khoo, Konar, Kong, Lester et~al.}}]{Barr:2011xt}
\bibinfo{author}{\bibfnamefont{A.}~\bibnamefont{Barr}},
  \bibinfo{author}{\bibfnamefont{T.}~\bibnamefont{Khoo}},
  \bibinfo{author}{\bibfnamefont{P.}~\bibnamefont{Konar}},
  \bibinfo{author}{\bibfnamefont{K.}~\bibnamefont{Kong}},
  \bibinfo{author}{\bibfnamefont{C.}~\bibnamefont{Lester}},
  \bibnamefont{et~al.}, \bibinfo{journal}{Phys.Rev.}
  \textbf{\bibinfo{volume}{D84}}, \bibinfo{pages}{095031}
  (\bibinfo{year}{2011}{\natexlab{a}}), \eprint{1105.2977}.

\bibitem[{\citenamefont{Barr and Gwenlan}(2009)}]{Barr:2009wu}
\bibinfo{author}{\bibfnamefont{A.~J.} \bibnamefont{Barr}} \bibnamefont{and}
  \bibinfo{author}{\bibfnamefont{C.}~\bibnamefont{Gwenlan}},
  \bibinfo{journal}{Phys.Rev.} \textbf{\bibinfo{volume}{D80}},
  \bibinfo{pages}{074007} (\bibinfo{year}{2009}), \eprint{0907.2713}.

\bibitem[{\citenamefont{Aad et~al.}(2011{\natexlab{a}})}]{daCosta:2011qk}
\bibinfo{author}{\bibfnamefont{G.}~\bibnamefont{Aad}} \bibnamefont{et~al.}
  (\bibinfo{collaboration}{ATLAS Collaboration}), \bibinfo{journal}{Phys.Lett.}
  \textbf{\bibinfo{volume}{B701}}, \bibinfo{pages}{186}
  (\bibinfo{year}{2011}{\natexlab{a}}), \eprint{1102.5290}.

\bibitem[{\citenamefont{Aad et~al.}(2012{\natexlab{b}})}]{Aad:2011ib}
\bibinfo{author}{\bibfnamefont{G.}~\bibnamefont{Aad}} \bibnamefont{et~al.}
  (\bibinfo{collaboration}{ATLAS Collaboration}), \bibinfo{journal}{Phys.Lett.}
  \textbf{\bibinfo{volume}{B710}}, \bibinfo{pages}{67}
  (\bibinfo{year}{2012}{\natexlab{b}}), \eprint{1109.6572}.

\bibitem[{\citenamefont{Aad et~al.}(2013)}]{Aad:2012pxa}
\bibinfo{author}{\bibfnamefont{G.}~\bibnamefont{Aad}} \bibnamefont{et~al.}
  (\bibinfo{collaboration}{ATLAS Collaboration}), \bibinfo{journal}{Phys.Lett.}
  \textbf{\bibinfo{volume}{B718}}, \bibinfo{pages}{879} (\bibinfo{year}{2013}),
  \eprint{1208.2884}.

\bibitem[{\citenamefont{Aad et~al.}(2012{\natexlab{c}})}]{Aad:2012uu}
\bibinfo{author}{\bibfnamefont{G.}~\bibnamefont{Aad}} \bibnamefont{et~al.}
  (\bibinfo{collaboration}{ATLAS Collaboration}), \bibinfo{journal}{JHEP}
  \textbf{\bibinfo{volume}{1211}}, \bibinfo{pages}{094}
  (\bibinfo{year}{2012}{\natexlab{c}}), \eprint{1209.4186}.

\bibitem[{ATL(2013)}]{ATL-PHYS-PUB-2013-004}
\bibinfo{type}{Tech. Rep.} \bibinfo{number}{ATL-PHYS-PUB-2013-004},
  \bibinfo{institution}{CERN}, \bibinfo{address}{Geneva}
  (\bibinfo{year}{2013}).

\bibitem[{\citenamefont{Cacciari et~al.}(2008)\citenamefont{Cacciari, Salam,
  and Soyez}}]{Cacciari:2008gp}
\bibinfo{author}{\bibfnamefont{M.}~\bibnamefont{Cacciari}},
  \bibinfo{author}{\bibfnamefont{G.~P.} \bibnamefont{Salam}}, \bibnamefont{and}
  \bibinfo{author}{\bibfnamefont{G.}~\bibnamefont{Soyez}},
  \bibinfo{journal}{JHEP} \textbf{\bibinfo{volume}{0804}}, \bibinfo{pages}{063}
  (\bibinfo{year}{2008}), \eprint{0802.1189}.

\bibitem[{\citenamefont{Cacciari et~al.}(2012)\citenamefont{Cacciari, Salam,
  and Soyez}}]{Cacciari:2011ma}
\bibinfo{author}{\bibfnamefont{M.}~\bibnamefont{Cacciari}},
  \bibinfo{author}{\bibfnamefont{G.~P.} \bibnamefont{Salam}}, \bibnamefont{and}
  \bibinfo{author}{\bibfnamefont{G.}~\bibnamefont{Soyez}},
  \bibinfo{journal}{Eur.Phys.J.} \textbf{\bibinfo{volume}{C72}},
  \bibinfo{pages}{1896} (\bibinfo{year}{2012}), \eprint{1111.6097}.

\bibitem[{\citenamefont{Arnold et~al.}(2009)\citenamefont{Arnold, Bahr, Bozzi,
  Campanario, Englert et~al.}}]{Arnold:2008rz}
\bibinfo{author}{\bibfnamefont{K.}~\bibnamefont{Arnold}},
  \bibinfo{author}{\bibfnamefont{M.}~\bibnamefont{Bahr}},
  \bibinfo{author}{\bibfnamefont{G.}~\bibnamefont{Bozzi}},
  \bibinfo{author}{\bibfnamefont{F.}~\bibnamefont{Campanario}},
  \bibinfo{author}{\bibfnamefont{C.}~\bibnamefont{Englert}},
  \bibnamefont{et~al.}, \bibinfo{journal}{Comput.Phys.Commun.}
  \textbf{\bibinfo{volume}{180}}, \bibinfo{pages}{1661} (\bibinfo{year}{2009}),
  \eprint{0811.4559}.

\bibitem[{\citenamefont{Hahn}(2001)}]{Hahn:2000kx}
\bibinfo{author}{\bibfnamefont{T.}~\bibnamefont{Hahn}},
  \bibinfo{journal}{Comput.Phys.Commun.} \textbf{\bibinfo{volume}{140}},
  \bibinfo{pages}{418} (\bibinfo{year}{2001}), \eprint{hep-ph/0012260}.

\bibitem[{\citenamefont{Hahn and Perez-Victoria}(1999)}]{Hahn:1998yk}
\bibinfo{author}{\bibfnamefont{T.}~\bibnamefont{Hahn}} \bibnamefont{and}
  \bibinfo{author}{\bibfnamefont{M.}~\bibnamefont{Perez-Victoria}},
  \bibinfo{journal}{Comput.Phys.Commun.} \textbf{\bibinfo{volume}{118}},
  \bibinfo{pages}{153} (\bibinfo{year}{1999}), \eprint{hep-ph/9807565}.

\bibitem[{\citenamefont{Boos et~al.}(2001)\citenamefont{Boos, Dobbs, Giele,
  Hinchliffe, Huston et~al.}}]{Boos:2001cv}
\bibinfo{author}{\bibfnamefont{E.}~\bibnamefont{Boos}},
  \bibinfo{author}{\bibfnamefont{M.}~\bibnamefont{Dobbs}},
  \bibinfo{author}{\bibfnamefont{W.}~\bibnamefont{Giele}},
  \bibinfo{author}{\bibfnamefont{I.}~\bibnamefont{Hinchliffe}},
  \bibinfo{author}{\bibfnamefont{J.}~\bibnamefont{Huston}},
  \bibnamefont{et~al.} (\bibinfo{year}{2001}), \eprint{hep-ph/0109068}.

\bibitem[{\citenamefont{Bahr et~al.}(2008)\citenamefont{Bahr, Gieseke, Gigg,
  Grellscheid, Hamilton et~al.}}]{Bahr:2008pv}
\bibinfo{author}{\bibfnamefont{M.}~\bibnamefont{Bahr}},
  \bibinfo{author}{\bibfnamefont{S.}~\bibnamefont{Gieseke}},
  \bibinfo{author}{\bibfnamefont{M.}~\bibnamefont{Gigg}},
  \bibinfo{author}{\bibfnamefont{D.}~\bibnamefont{Grellscheid}},
  \bibinfo{author}{\bibfnamefont{K.}~\bibnamefont{Hamilton}},
  \bibnamefont{et~al.}, \bibinfo{journal}{Eur.Phys.J.}
  \textbf{\bibinfo{volume}{C58}}, \bibinfo{pages}{639} (\bibinfo{year}{2008}),
  \eprint{0803.0883}.

\bibitem[{\citenamefont{Dawson et~al.}(1998)\citenamefont{Dawson, Dittmaier,
  and Spira}}]{Dawson:1998py}
\bibinfo{author}{\bibfnamefont{S.}~\bibnamefont{Dawson}},
  \bibinfo{author}{\bibfnamefont{S.}~\bibnamefont{Dittmaier}},
  \bibnamefont{and} \bibinfo{author}{\bibfnamefont{M.}~\bibnamefont{Spira}},
  \bibinfo{journal}{Phys.Rev.} \textbf{\bibinfo{volume}{D58}},
  \bibinfo{pages}{115012} (\bibinfo{year}{1998}), \eprint{hep-ph/9805244}.

\bibitem[{\citenamefont{Gleisberg et~al.}(2009)\citenamefont{Gleisberg, Hoeche,
  Krauss, Schonherr, Schumann et~al.}}]{Gleisberg:2008ta}
\bibinfo{author}{\bibfnamefont{T.}~\bibnamefont{Gleisberg}},
  \bibinfo{author}{\bibfnamefont{S.}~\bibnamefont{Hoeche}},
  \bibinfo{author}{\bibfnamefont{F.}~\bibnamefont{Krauss}},
  \bibinfo{author}{\bibfnamefont{M.}~\bibnamefont{Schonherr}},
  \bibinfo{author}{\bibfnamefont{S.}~\bibnamefont{Schumann}},
  \bibnamefont{et~al.}, \bibinfo{journal}{JHEP}
  \textbf{\bibinfo{volume}{0902}}, \bibinfo{pages}{007} (\bibinfo{year}{2009}),
  \eprint{0811.4622}.

\bibitem[{\citenamefont{Alwall et~al.}(2011)\citenamefont{Alwall, Herquet,
  Maltoni, Mattelaer, and Stelzer}}]{Alwall:2011uj}
\bibinfo{author}{\bibfnamefont{J.}~\bibnamefont{Alwall}},
  \bibinfo{author}{\bibfnamefont{M.}~\bibnamefont{Herquet}},
  \bibinfo{author}{\bibfnamefont{F.}~\bibnamefont{Maltoni}},
  \bibinfo{author}{\bibfnamefont{O.}~\bibnamefont{Mattelaer}},
  \bibnamefont{and} \bibinfo{author}{\bibfnamefont{T.}~\bibnamefont{Stelzer}},
  \bibinfo{journal}{JHEP} \textbf{\bibinfo{volume}{1106}}, \bibinfo{pages}{128}
  (\bibinfo{year}{2011}), \eprint{1106.0522}.

\bibitem[{\citenamefont{Denner et~al.}(2011)\citenamefont{Denner, Dittmaier,
  Kallweit, and Pozzorini}}]{Denner:2010jp}
\bibinfo{author}{\bibfnamefont{A.}~\bibnamefont{Denner}},
  \bibinfo{author}{\bibfnamefont{S.}~\bibnamefont{Dittmaier}},
  \bibinfo{author}{\bibfnamefont{S.}~\bibnamefont{Kallweit}}, \bibnamefont{and}
  \bibinfo{author}{\bibfnamefont{S.}~\bibnamefont{Pozzorini}},
  \bibinfo{journal}{Phys.Rev.Lett.} \textbf{\bibinfo{volume}{106}},
  \bibinfo{pages}{052001} (\bibinfo{year}{2011}), \eprint{1012.3975}.

\bibitem[{\citenamefont{Campbell and Ellis}(2000)}]{zbb1}
\bibinfo{author}{\bibfnamefont{J.~M.} \bibnamefont{Campbell}} \bibnamefont{and}
  \bibinfo{author}{\bibfnamefont{R.~K.} \bibnamefont{Ellis}},
  \bibinfo{journal}{Phys.Rev.} \textbf{\bibinfo{volume}{D62}},
  \bibinfo{pages}{114012} (\bibinfo{year}{2000}), \eprint{hep-ph/0006304}.

\bibitem[{\citenamefont{Campbell}(2001)}]{zbb2}
\bibinfo{author}{\bibfnamefont{J.~M.} \bibnamefont{Campbell}}
  (\bibinfo{year}{2001}), \eprint{hep-ph/0105226}.

\bibitem[{\citenamefont{Campbell et~al.}(2011)\citenamefont{Campbell, Ellis,
  and Williams}}]{Campbell:2011bn}
\bibinfo{author}{\bibfnamefont{J.~M.} \bibnamefont{Campbell}},
  \bibinfo{author}{\bibfnamefont{R.~K.} \bibnamefont{Ellis}}, \bibnamefont{and}
  \bibinfo{author}{\bibfnamefont{C.}~\bibnamefont{Williams}},
  \bibinfo{journal}{JHEP} \textbf{\bibinfo{volume}{1107}}, \bibinfo{pages}{018}
  (\bibinfo{year}{2011}), \eprint{1105.0020}.

\bibitem[{\citenamefont{Chatrchyan
  et~al.}(2012{\natexlab{b}})}]{Chatrchyan:2012vp}
\bibinfo{author}{\bibfnamefont{S.}~\bibnamefont{Chatrchyan}}
  \bibnamefont{et~al.} (\bibinfo{collaboration}{CMS Collaboration}),
  \bibinfo{journal}{Phys.Lett.} \textbf{\bibinfo{volume}{B713}},
  \bibinfo{pages}{68} (\bibinfo{year}{2012}{\natexlab{b}}), \eprint{1202.4083}.

\bibitem[{\citenamefont{Aad et~al.}(2012{\natexlab{d}})}]{Aad:2012mea}
\bibinfo{author}{\bibfnamefont{G.}~\bibnamefont{Aad}} \bibnamefont{et~al.}
  (\bibinfo{collaboration}{ATLAS Collaboration}), \bibinfo{journal}{JHEP}
  \textbf{\bibinfo{volume}{1209}}, \bibinfo{pages}{070}
  (\bibinfo{year}{2012}{\natexlab{d}}), \eprint{1206.5971}.

\bibitem[{\citenamefont{Barr et~al.}(2011{\natexlab{b}})\citenamefont{Barr,
  French, Frost, and Lester}}]{Barr:2011he}
\bibinfo{author}{\bibfnamefont{A.~J.} \bibnamefont{Barr}},
  \bibinfo{author}{\bibfnamefont{S.~T.} \bibnamefont{French}},
  \bibinfo{author}{\bibfnamefont{J.~A.} \bibnamefont{Frost}}, \bibnamefont{and}
  \bibinfo{author}{\bibfnamefont{C.~G.} \bibnamefont{Lester}},
  \bibinfo{journal}{JHEP} \textbf{\bibinfo{volume}{1110}}, \bibinfo{pages}{080}
  (\bibinfo{year}{2011}{\natexlab{b}}), \eprint{1106.2322}.

\bibitem[{\citenamefont{Aad et~al.}(2011{\natexlab{b}})}]{taurec1}
\bibinfo{author}{\bibfnamefont{G.}~\bibnamefont{Aad}} \bibnamefont{et~al.}
  (\bibinfo{collaboration}{ATLAS Collaboration}), \bibinfo{journal}{Phys.Rev.}
  \textbf{\bibinfo{volume}{D84}}, \bibinfo{pages}{112006}
  (\bibinfo{year}{2011}{\natexlab{b}}), \eprint{1108.2016}.

\bibitem[{\citenamefont{Elagin et~al.}(2011)\citenamefont{Elagin, Murat,
  Pranko, and Safonov}}]{taurec2}
\bibinfo{author}{\bibfnamefont{A.}~\bibnamefont{Elagin}},
  \bibinfo{author}{\bibfnamefont{P.}~\bibnamefont{Murat}},
  \bibinfo{author}{\bibfnamefont{A.}~\bibnamefont{Pranko}}, \bibnamefont{and}
  \bibinfo{author}{\bibfnamefont{A.}~\bibnamefont{Safonov}},
  \bibinfo{journal}{Nucl.Instrum.Meth.} \textbf{\bibinfo{volume}{A654}},
  \bibinfo{pages}{481} (\bibinfo{year}{2011}), \eprint{1012.4686}.

\bibitem[{\citenamefont{Edwards}(1972)}]{edwards}
\bibinfo{author}{\bibfnamefont{A.~W.~F.} \bibnamefont{Edwards}},
  \emph{\bibinfo{title}{Likelihoods}} (\bibinfo{publisher}{Cambridge University
  Press}, \bibinfo{year}{1972}).

\bibitem[{\citenamefont{Read}(2002)}]{Read:2002hq}
\bibinfo{author}{\bibfnamefont{A.~L.} \bibnamefont{Read}},
  \bibinfo{journal}{J.Phys.} \textbf{\bibinfo{volume}{G28}},
  \bibinfo{pages}{2693} (\bibinfo{year}{2002}).

\bibitem[{\citenamefont{Plehn et~al.}(2011)\citenamefont{Plehn, Spannowsky, and
  Takeuchi}}]{Plehn:2011tf}
\bibinfo{author}{\bibfnamefont{T.}~\bibnamefont{Plehn}},
  \bibinfo{author}{\bibfnamefont{M.}~\bibnamefont{Spannowsky}},
  \bibnamefont{and} \bibinfo{author}{\bibfnamefont{M.}~\bibnamefont{Takeuchi}},
  \bibinfo{journal}{JHEP} \textbf{\bibinfo{volume}{1105}}, \bibinfo{pages}{135}
  (\bibinfo{year}{2011}), \eprint{1102.0557}.

\end{thebibliography}
\end{document}